\documentclass[aps,pre,showkeys,showpacs,reprint]{revtex4-1}

\usepackage{natbib,graphicx,amssymb,amsmath,xcolor,bm,hyperref}

\begin{document}
\title[]{Extra invariant and plasma inhomogeneity to improve zonal flow}

\author{Alexander M. Balk \email[balk@math.utah.edu]}
\affiliation{Department of Mathematics, University of Utah,
155 South 1400 East, Salt Lake City, Utah 84112}

\date{\today}

\begin{abstract} 
Zonal flows are known to diminish turbulent transport in magnetic fusion. Interstingly, there is an adiabatic invariant that implies the emergence of zonal flow. The paper shows that if this invariant is decreasing (due to some external factors) then the emerging zonal flow is better. It is also shown that the plasma inhomogeneity can lead to the decrease of the adiabatic invariant. A simple condition for such decrease is found.
 \end{abstract} 

\keywords{Fusion plasma, Transport barriers, drift/Rossby waves, Additional conservation.}
\maketitle

\section{Introduction}
\label{Intro}

One of the problems of nuclear fusion is to decrease transport of heat and particles from the core of a fusion device to its walls. In devices with magnetic confinement, the decreased transport could be achieved by creating strong alternating poloidal flow (often called ``zonal'' flow in analogy with hydrodynamics). This is reviewed in \cite{Horton99,Diamond}. The possibility to decrease transport is based on the idea  \cite{BiglariDiTe} that alternating zonal/poloidal flow interrupts ballistic transport (decreases mean free path) across the flow. At the same time, the poloidal flow does not contribute to the radial transport. 

We consider zonal/poloidal flow not as the flow with exactly zero poloidal wave number ($q=0$), but as the turbulence whose poloidal wave number is much less than the radial wave number ($|q|\ll |p|$). The zonal flow turbulence is a continuous continuation of other plasma turbulence with the entire spectrum of wave vectors ${\bf k}=(p,q)$. So, zonal flow that efficiently reduces transport, should have a large amount of large-scale energy tightly concentrated around the $p$-axis. This energy will eventually lead to the coherent zonal flow with exactly zero poloidal number (like the emergence of large vortex in 2D hydrodynamics), but this process is not considered in the present paper. The problem of interaction between the drift turbulence and the coherent zonal flow was studied for decades and is intensively studied now, which is reviewed in \cite{MarstonTobias}.

We suppose that a significant mode of plasma dynamics represent drift waves with dispersion relation
\begin{eqnarray}\label{DispLaw}
\omega({\bf k})=\frac{\nu+\mu k^2}{1+k^2}\,q,
\end{eqnarray}
where the wave vector ${\bf k}=(p,q)$ is non-dimensionalized by the Larmor (Rossby) radius $\rho$, and $k^2=p^2+q^2$.
The interaction of these waves can have different forms, and we do not need to specify it.
The frequencies $\mu$ and $\nu$ can depend on  the coherent part of the poloidal/zonal flow \cite{parker16,Ruiz,ZhuDodin}. More about this is in  Section \ref{Kinetic}. 

The waves similar to the plasma drift waves occur in different problems. As well known, the waves with the same dispersion relation are the Rossby waves in atmosphere and ocean \cite{Va}; then often $\nu=\beta\rho$ ($\beta$ is the Coriolis parameter) and $\mu=0$. The dynamics of Rossby waves leads to the generation of zonal flows in atmospheres and oceans on several planets. An example of this phenomenon are Jupiter's stripes. Much slower waves (slower than the relatively slow Rossby waves) also appear in astrophysical magnethydrodynamics \cite{Hide66,ZaqOliBalShe,Balk2014,B2022}. Such waves are often called the magnetic Rossby waves. In this case, $\nu=0$ and $\mu \rho^3=B_0^2/\beta$ ($B_0$ is the basic zonal magnetic field, measured in the velocity units). These waves occur in the upper layer of Earth's liquid iron outer core (``ocean of the core'' \cite{Brag98}). They also take place in Sun's tachocline.

It is interesting that the emergence of zonal/poloidal flow follows from the extra conservation. To describe this extra invariant, let us first recall the question considered by L. Boltzmann \cite{Boltz1,Boltz2} about binary collisions
\begin{subequations}\label{Boltz}
\begin{eqnarray}
{\bf p}_1+{\bf p}_2&=&{\bf p}_3+{\bf p}_4,\\
{\bf p}_1^2+{\bf p}_2^2&=&{\bf p}_3^2+{\bf p}_4^2
\end{eqnarray}
(${\bf p}_i$ are the momentum vectors of the two particles before and after the collision), see also \cite{Boltzmann,BaFe}. Does there exist an independent quantity $\phi({\bf p})$ which is conserved in the collisions (\ref{Boltz}ab) 
\begin{eqnarray}
\phi({\bf p}_1)+\phi({\bf p}_2)=\phi({\bf p}_3)+\phi({\bf p}_4) ?
\end{eqnarray}
In other words, does there exist a function  $\phi({\bf p})$ such that the equation  (\ref{Boltz}c) holds for any vectors ${\bf p}_i\; (i=1,2,3,4)$ bound by the relations (\ref{Boltz}ab)? And the function $\phi({\bf p})$ is not a mere  linear combination of the momentum ${\bf p}$ and energy ${\bf p}^2$.
\end{subequations}
Boltzmann had shown that such quantity does not exist. 

Similar, one can consider resonance interactions of drift/Rossby waves
\begin{subequations}\label{Rossby}
\begin{eqnarray}
{\bf k}_1&=&{\bf k}_2+{\bf k}_3,\\
\omega({\bf k}_1)&=&\omega({\bf k}_2)+\omega({\bf k}_3)
\end{eqnarray}
(${\bf k}_i$ are the wave vectors) and pose the question: Does there exist an independent quantity $\eta({\bf k})$ which is conserved in the interactions (\ref{Rossby}ab) 
\begin{eqnarray}
\eta({\bf k}_1)=\eta({\bf k}_2)+\eta({\bf k}_3) ?
\end{eqnarray}
In other words, does there exist a function  $\eta({\bf k})$ such that the equation  (\ref{Rossby}c) holds for any vectors ${\bf k}_i\; (i=1,2,3)$ bound by the relations (\ref{Rossby}ab)? And the function $\eta({\bf k})$ is not a mere  linear combination of the momentum ${\bf k}$ and energy $\omega({\bf k})$.
\end{subequations}
For the dispersion relation (\ref{DispLaw}) such function does exist, namely \cite{BNZ,B1991}
\begin{eqnarray}\label{extraDnsty}
\eta({\bf k})=\arctan{\frac{p+q\sqrt{3}}{k^2}}-\arctan{\frac{p-q\sqrt{3}}{k^2}}\,.
\end{eqnarray}
Thus, in addition to the energy and momentum, a system of drift/Rossby waves has an extra invariant
${\mathcal I}=\int \eta_{\bf k} {N}_{\bf k}\, d{\bf k}$. 
The function ${N}_{\bf k}=\varepsilon_{\bf k}/\omega_{\bf k}$ is the wave action spectrum (or the phase space density of quasi-particles), $\varepsilon_{\bf k}$ is the energy spectrum. 
(To have positive ${N}_{\bf k}$, one needs to restrict wave vectors ${\bf k}$ to half-plane $q>0$.)

The extra invariant stems from the integrability work by V. Zakharov and E. Shulman, published in a series of papers throughout 1980s (in particular, \cite{ZSch0,ZSch}). They found that conservation in resonance interactions implies cancellation of  small divisors, and therefore, leads to an additional approximate invariant of the corresponding Hamiltonian system. An important feature of their theory is that one can make physical predictions without knowing the exact form of nonlinearity. It is only required that the wave dynamics is Hamiltonian (which is usually the case for a physical system). Some aspects of this theory turned out \cite{Ferapontov} to be related to the web geometry \cite{Bl1}. This allowed to show \cite{BaFe} that the Rossby wave extra invariant $\mathcal I$ (or the extra conserved quantity $\eta$) is {\it unique:} All  invariants are linear combinations of the energy, the enstrophy, and the extra invariant $\mathcal I$.

The extra conservation implies the emergence of zonal flow \cite{B2005}, see also Section \ref{Decreasing}. Here we will see even more: If the extra invariant is decreasing with time (due to some external factors), then the emerging zonal flow is better; it diminishes transport even more efficiently. 

The extra invariant and its physical implications were established through the work of several people, in particular, \cite{Glp,Weichman06,BaVan,SukorianskyDG,LeeSmith,Read+6,Yoshi-Yuki07,BaYo,Nazarenko,BaZaSimilarity,
BHW,Weichman12,SaitoIshioka13,SaitoIshioka16}. 
In \cite{Nazarenko}, the adiabatic invariant $I$ was called ``zonostrophy'' (``zonostrophic instability'' is unrelated).

\section{Decreasing extra invariant leads to efficient zonal flow.}
\label{Decreasing}
First, let us note the energy accumulation at large scales. This is obvious for Rossby waves when $\mu=0$ and $\nu\neq0$; then the energy follows the inverse cascade. But even in general, the energy accumulation at large scales appears to be true. For instance, in the classic 3-dimensional hydrodynamic turbulence, the energy follows the direct cascade, but the large scales carry most of the energy (the Kolmogorov spectrum has infrared divergence). Similar situation occurs for the gravity waves on the ocean surface \cite{ZakhKS}. This system conserves two integrals: the energy and the wave action. The wave action follows the inverse cascade, while the energy follows the direct cascade. Still, most of the energy is carried by large scales (the Kolmogorov-Zakharov energy spectrum, determined by the wave action flux, has infrared divergence). In both of these situations, most of the energy flows towards large $k$ and dissipates there. However, a small fraction of the energy reaches large scales, and the energy accumulates there.
Similar phenomenon seems to hold \cite{Balk2014,B2022} for the turbulence of magnetic Rossby waves, when $\nu=0$ and $\mu\neq0$. Then the energy and the enstrophy trade places; and so, there is direct cascade of energy, but the energy accumulates at small $k$. When both parameters $\mu$ and $\nu$ in (\ref{DispLaw}) are non-zero, then there is no clear direction for the energy cascade, but presumably, the energy accumulates at large scales. 

\begin{figure}
\includegraphics[width=\columnwidth]{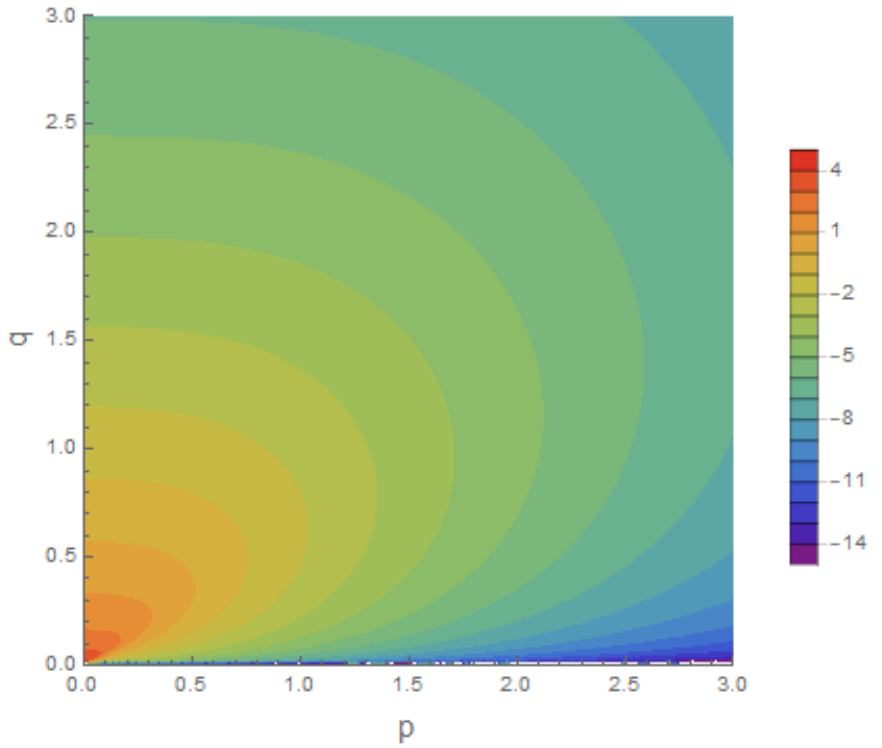}
\caption{Contour plot of the ratio $R_{\bf k}=\tilde{\eta}_{\bf k}/\omega_{\bf k}$ in logarithmic scale: The color represents the values of $\ln(R_{\bf k})$. The figure is qualitatively the same for various value of the parameters $\mu$ and $\nu$ in (\ref{DispLaw}); for this particular plot, $\mu=1,\nu=2$. The graph is shown only for positive $p,q$, due to the symmetries $R(p,q)=R(-p,q)=R(p,-q)$.}
\label{fig:ratio}
\end{figure}

Now recall the well known conserved quantities for the system of interacting drift waves. These are the energy $\mathcal E$ and the zonal momentum $\mathcal M$
\begin{eqnarray}\label{EM}
\mathcal E=\int\omega_{\bf k}\; N_{\bf k}\; d{\bf k},\qquad
\mathcal M=\int q\; N_{\bf k}\; d{\bf k}.
\end{eqnarray}
The enstrophy is a linear combination of $\mathcal E$ and $\mathcal M$. The $p$-momentum is not a real physical invariant because the function $p/\omega_{\bf k}$ has singularity (when $q\rightarrow0$), see \cite{BaYo} for details.
\begin{subequations}\label{mI}
To show the emergence of zonal flow, we need to consider the following linear combination of the extra invariant $\mathcal I$ with $\mathcal E$ and $\mathcal M$
\begin{eqnarray}
\tilde{\mathcal I}=\mathcal I-2\sqrt{3}\,\frac{\mathcal E-\mu\, \mathcal M}{\nu-\mu}=\int \tilde{\eta}_{\bf k}\, N_{\bf k}\, d{\bf k},
\end{eqnarray}
where 
\begin{eqnarray}
\tilde{\eta}_{\bf k}=\eta_{\bf k}-2\sqrt{3}\,\frac{\omega_{\bf k}-\mu\, q}{\nu-\mu}=
\eta_{\bf k}-2\sqrt{3}\,\frac{p}{1+k^2}. 
\end{eqnarray}
Like $\eta_{\bf k}$, the function $\tilde{\eta}_{\bf k}$ is independent of $\mu$ and $\nu$.
\end{subequations}
Let us write the modified extra invariant (\ref{mI}) in terms of the energy spectrum: 
$\tilde{\mathcal I}=\int R_{\bf k}\,\varepsilon_{\bf k}\; d{\bf k}$, 
where  the ratio $R_{\bf k}=\tilde{\eta}_{\bf k}/\omega_{\bf k}$
shows how much extra invariant $\tilde{\mathcal I}$ is carried by the unit amount of energy with wave vector 
${\bf k}=(p,q)$. The contour plot of $R_{\bf k}$ is presented in figure \ref{fig:ratio}. 

Figure \ref{fig:ratio} shows that the accumulated large scale energy should concentrate near the $p$-axis (poloidal/zonal flow). Indeed, if during the energy transfer towards large scales, the energy accumulates away from $p$-axis, then the extra invariant $\tilde{\mathcal I}$ would significantly increase (in contradiction to the extra conservation). This is because --- according to the figure ---  $R_{\bf k}$ is small at large $k$ but large at small $k$, unless $q$ is close to zero. This is also supported by the asymptotic behavior:
\begin{eqnarray}\label{asy}
\frac{(\nu+\mu k^2)R_{\bf k}}{8\sqrt{3}}=\left\{\begin{array}{ll}
\frac{q^2 (5p^2+q^2)}{5 k^8}+O(k^{-6}) & (k\rightarrow\infty),\\ \\
\frac{q^2}{p^2 (1+p^2)^2}+O(q^4) & (q\rightarrow0).\end{array}\right.\quad
\end{eqnarray}

Now suppose, the extra invariant $\tilde{I}$ is decreasing because of some external factors, but the energy $\mathcal E$ is not decreasing. Then a lesser amount of the extra invariant is available for the spreading of the energy away from the $p$-axis, and the energy has to concentrate more tightly around $p$-axis. This means that the emerging zonal flow is  more unidirectional, and it contributes less to the radial transport. In addition, the zonal flow interrupts the radial transport, decreasing the mean free path. It is worthwhile to note that the transport decrease occurs for transport by any plasma mode (not just by drift waves). 

Actually, one needs to decrease the extra invariant as much as possible, without decreasing the energy. This is relatively easy because the energy kernel (\ref{DispLaw}) and the extra invariant kernel (\ref{extraDnsty}) are quite different.

There are at least two ways to decrease the extra invariant without decreasing the energy.

First, one can introduce some dissipation and pumping in different areas of the ${\bf k}$-plane, so that the extra invariant $\tilde{\mathcal I}$ is dissipated, while the energy $\mathcal E$ is pumped. This approach was studied in \cite{B19}. 

The present paper reports another opportunity which is due to the inhomogeneity of fusion plasma. 

\section{Evolution of the invariant by the inhomogeneous wave kinetic equation}
\label{Kinetic}
Let us describe the turbulence of the drift/Rossby waves by the wave action spectrum $N(p,q,x,y,t)$ obeying the wave kinetic equation 
\begin{eqnarray}\label{WKE}
\frac{\partial {N}}{\partial t}+
  \frac{\partial (\omega,N)}{\partial ({\bf k}, {\bf x})}=St[{N}]\,,
\end{eqnarray}
where we use the notation
\begin{eqnarray*}
\frac{\partial ({\mathcal F},{\mathcal G})}{\partial(u,v)}=
\frac{\partial {\mathcal F}}{\partial u}\cdot \frac{\partial {\mathcal G}}{\partial v}\;-\;
\frac{\partial {\mathcal F}}{\partial v}\cdot \frac{\partial {\mathcal G}}{\partial u}
\end{eqnarray*}    
for arbitrary functions ${\mathcal F}$, ${\mathcal G}$ of arbitrary variables $u$, $v$ (vector or scalar). 
The l.h.s.\ in equation (\ref{WKE}) describes the slow refraction of waves, while the r.h.s. (stoss-term or collision integral) describes changes in $N$ due to the 3-wave resonance interactions (\ref{Rossby}).
We assume that the interactions occur locally in physical space, and so, the three interacting waves have the same plasma parameters (the inhomogeneity does not enter $St[N]$).

Let us consider the evolution by the equation (\ref{WKE}) of the integral
\begin{eqnarray}\label{InhomoInvar}
\tilde{\mathcal I}=\int\tilde{\eta}(p, q) \; N(p,q,x,y,t)\; dp\,dq.
\end{eqnarray}
We multiply equation (\ref{WKE}) by $\tilde\eta$ and integrate over $d{\bf k}$. Since the quantity $\tilde\eta$ is conserved in the 3-wave resonance interactions, the term $St[N]\tilde\eta$ integrates to zero. The l.h.s. in (\ref{WKE}) leads to the flowing of $\tilde I$ in the physical plane (the divergence of the flux) and to a source/sink term
\begin{eqnarray}\label{D}
\frac{\partial {\tilde{\mathcal I}}}{\partial t}+
\frac{\partial}{\partial {\bf x}}\cdot\int\frac{\partial\omega}{\partial {\bf k}}\,{\tilde\eta}\,N\,d{\bf k}=
\int S\, N\,d{\bf k}
\end{eqnarray}
with the kernel
\begin{eqnarray}\label{S}
&&S=\frac{\partial (\omega,{\tilde\eta})}{\partial({\bf k},{\bf x})}=
-\frac{\partial\omega}{\partial x}\frac{\partial{\tilde \eta}}{\partial p}= \nonumber\\
&&\frac{p\,(\nu'+k^2\mu')\;\; 16\sqrt{3}\,(1+4p^2)\,q^4}{(1+k^2)^3[(p+\sqrt{3}q)^2+k^4][(p-\sqrt{3}q)^2+k^4]}\,,
\end{eqnarray}
where the frequencies $\mu$ and $\nu$ in the dispersion relation (\ref{DispLaw}) slowly depend on the radial coordinate $x$; prime denotes the $x$-derivative.

If the function $\tilde\eta$ were replaced by $\omega$, then the last integral in (\ref{D}) would vanish, and we would have the energy conservation (energy flows throughout the $xy$-plane). So, the inhomogeneous part of the kinetic equation can decrease the extra invariant $\tilde{\mathcal I}$, while the energy $\mathcal E$ is always conserved by the entire kinetic equation.

The parameters of the dispersion relation (\ref{DispLaw}) depend on the  zonal flow velocity $U(x)$ \cite{parker16,Ruiz,ZhuDodin}
\begin{eqnarray}\label{MuNu}
\mu=U(x)/\rho,\qquad\nu=\rho[\kappa  +U''(x)]+U(x)/\rho\,.\quad
\end{eqnarray}
The radial $x$ and poloidal $y$ coordinates are considered in generalized sense: Radial direction is the direction of the anti-gradient of some quantity $Q$, the poloidal direction is orthogonal to the radial direction and to the confining magnetic field (the contour lines $Q=$const are not necessarily circles). The constant $\kappa$ is determined by the local value of the gradient $\nabla Q$ ($\kappa$ is analogous to the $\beta$-parameter in hydrodynamics). 

The $U''$ term in the second expression (\ref{MuNu}) is often neglected, but recently its significance is re-considered \cite{parker16,Ruiz,ZhuDodin}.

We see from (\ref{D})-(\ref{S}) that the invariant $\tilde I$ would decrease if the drift wave turbulence spectrum had significant asymmetric part
\begin{eqnarray}\label{Na}
N_a(p,q,x,y,t)=N(p,q,x.y,t)-N(-p,q,x,y,t) \nonumber
\end{eqnarray}
and
\begin{eqnarray}\label{cond}
[\rho^2 U'''+(1+k^2) U'] N_a<0
\end{eqnarray}
(most of the factors in (\ref{S}) are positive and cancel out). The condition (\ref{cond}) is especially simple if the $U''$ term in (\ref{MuNu}) is neglected. Then $\tilde I$ decreases when $U' N_a<0$. This means: If zonal flow velocity $U(x)$ is decreasing, the turbulence spectrum should be skewed towards positive $p$ (the spectrum $N$ should be bigger for positive $p$ than for negative $p$). If zonal flow velocity $U(x)$ is increasing, the turbulence spectrum should be skewed towards negative $p$. In general, the condition (\ref{cond}) involves the length scale. The condition (\ref{cond}) can be used to create a transport barrier at a certain location.

\section{Conclusion}
The dynamics of interacting drift waves has an adiabatic invariant; Section \ref{Intro}. The presence of this extra invariant implies the emergence of zonal flow, which diminishes the transport of heat and particles (by any plasma mode, not just by drift waves), and so, serves as transport barrier.
The present paper makes two points:

1). If the extra invariant is decreasing (due to some external factors), while the energy is not decreasing, then the emerging zonal flow is a better transport barrier; Section \ref{Decreasing}. The extra invariant kernel so much differs from the energy kernel, that it is relatively easy to decrease the extra invariant without decreasing the energy.

2). The plasma inhomogeneity can decrease the extra invariant, while preserving the energy; Section \ref{Kinetic}. To see this, we have applied the inhomogeneous wave kinetic equation (the three waves involved in the collision integral having the same plasma parameters) and derived a simple condition (\ref{cond}) for the decrease of the extra invariant.

\bibliography{My}

\begin{thebibliography}{40}%
\makeatletter
\providecommand \@ifxundefined [1]{%
 \@ifx{#1\undefined}
}%
\providecommand \@ifnum [1]{%
 \ifnum #1\expandafter \@firstoftwo
 \else \expandafter \@secondoftwo
 \fi
}%
\providecommand \@ifx [1]{%
 \ifx #1\expandafter \@firstoftwo
 \else \expandafter \@secondoftwo
 \fi
}%
\providecommand \natexlab [1]{#1}%
\providecommand \enquote  [1]{``#1''}%
\providecommand \bibnamefont  [1]{#1}%
\providecommand \bibfnamefont [1]{#1}%
\providecommand \citenamefont [1]{#1}%
\providecommand \href@noop [0]{\@secondoftwo}%
\providecommand \href [0]{\begingroup \@sanitize@url \@href}%
\providecommand \@href[1]{\@@startlink{#1}\@@href}%
\providecommand \@@href[1]{\endgroup#1\@@endlink}%
\providecommand \@sanitize@url [0]{\catcode `\\12\catcode `\$12\catcode
  `\&12\catcode `\#12\catcode `\^12\catcode `\_12\catcode `\%12\relax}%
\providecommand \@@startlink[1]{}%
\providecommand \@@endlink[0]{}%
\providecommand \url  [0]{\begingroup\@sanitize@url \@url }%
\providecommand \@url [1]{\endgroup\@href {#1}{\urlprefix }}%
\providecommand \urlprefix  [0]{URL }%
\providecommand \Eprint [0]{\href }%
\providecommand \doibase [0]{http://dx.doi.org/}%
\providecommand \selectlanguage [0]{\@gobble}%
\providecommand \bibinfo  [0]{\@secondoftwo}%
\providecommand \bibfield  [0]{\@secondoftwo}%
\providecommand \translation [1]{[#1]}%
\providecommand \BibitemOpen [0]{}%
\providecommand \bibitemStop [0]{}%
\providecommand \bibitemNoStop [0]{.\EOS\space}%
\providecommand \EOS [0]{\spacefactor3000\relax}%
\providecommand \BibitemShut  [1]{\csname bibitem#1\endcsname}%
\let\auto@bib@innerbib\@empty
\bibitem [{\citenamefont {Horton}(1999)}]{Horton99}%
  \BibitemOpen
  \bibfield  {author} {\bibinfo {author} {\bibfnamefont {W.}~\bibnamefont
  {Horton}},\ }\href@noop {} {\bibfield  {journal} {\bibinfo  {journal}
  {Reviews of Modern Physics}\ }\textbf {\bibinfo {volume} {71}},\ \bibinfo
  {pages} {735} (\bibinfo {year} {1999})}\BibitemShut {NoStop}%
\bibitem [{\citenamefont {Diamond}\ \emph {et~al.}(2005)\citenamefont
  {Diamond}, \citenamefont {Itoh}, \citenamefont {Itoh},\ and\ \citenamefont
  {Hahm}}]{Diamond}%
  \BibitemOpen
  \bibfield  {author} {\bibinfo {author} {\bibfnamefont {P.~H.}\ \bibnamefont
  {Diamond}}, \bibinfo {author} {\bibfnamefont {S.-I.}\ \bibnamefont {Itoh}},
  \bibinfo {author} {\bibfnamefont {K.}~\bibnamefont {Itoh}}, \ and\ \bibinfo
  {author} {\bibfnamefont {T.~S.}\ \bibnamefont {Hahm}},\ }\href@noop {}
  {\bibfield  {journal} {\bibinfo  {journal} {Plasma Phys. Control. Fusion}\
  }\textbf {\bibinfo {volume} {47}},\ \bibinfo {pages} {R35} (\bibinfo {year}
  {2005})}\BibitemShut {NoStop}%
\bibitem [{\citenamefont {Biglari}\ \emph {et~al.}(1990)\citenamefont
  {Biglari}, \citenamefont {Diamond},\ and\ \citenamefont
  {Terry}}]{BiglariDiTe}%
  \BibitemOpen
  \bibfield  {author} {\bibinfo {author} {\bibfnamefont {H.}~\bibnamefont
  {Biglari}}, \bibinfo {author} {\bibfnamefont {P.}~\bibnamefont {Diamond}}, \
  and\ \bibinfo {author} {\bibfnamefont {P.}~\bibnamefont {Terry}},\
  }\href@noop {} {\bibfield  {journal} {\bibinfo  {journal} {Physics of Fluids
  B: Plasma Physics}\ }\textbf {\bibinfo {volume} {2}},\ \bibinfo {pages} {1}
  (\bibinfo {year} {1990})}\BibitemShut {NoStop}%
\bibitem [{\citenamefont {Marston}\ and\ \citenamefont
  {Tobias}(2023)}]{MarstonTobias}%
  \BibitemOpen
  \bibfield  {author} {\bibinfo {author} {\bibfnamefont {J.}~\bibnamefont
  {Marston}}\ and\ \bibinfo {author} {\bibfnamefont {S.}~\bibnamefont
  {Tobias}},\ }\href {\doibase 10.1146/annurev-fluid-120720-031006} {\bibfield
  {journal} {\bibinfo  {journal} {Annual Review of Fluid Mechanics}\ }\textbf
  {\bibinfo {volume} {55}},\ \bibinfo {pages} {351} (\bibinfo {year} {2023})},\
  \Eprint
  {http://arxiv.org/abs/https://doi.org/10.1146/annurev-fluid-120720-031006}
  {https://doi.org/10.1146/annurev-fluid-120720-031006} \BibitemShut {NoStop}%
\bibitem [{\citenamefont {Parker}(2016)}]{parker16}%
  \BibitemOpen
  \bibfield  {author} {\bibinfo {author} {\bibfnamefont {J.~B.}\ \bibnamefont
  {Parker}},\ }\href {\doibase 10.1017/S0022377816001021} {\bibfield  {journal}
  {\bibinfo  {journal} {Journal of Plasma Physics}\ }\textbf {\bibinfo {volume}
  {82}},\ \bibinfo {pages} {595820602} (\bibinfo {year} {2016})}\BibitemShut
  {NoStop}%
\bibitem [{\citenamefont {Ruiz}\ \emph {et~al.}(2019)\citenamefont {Ruiz},
  \citenamefont {Glinsky},\ and\ \citenamefont {Dodin}}]{Ruiz}%
  \BibitemOpen
  \bibfield  {author} {\bibinfo {author} {\bibfnamefont {D.}~\bibnamefont
  {Ruiz}}, \bibinfo {author} {\bibfnamefont {M.}~\bibnamefont {Glinsky}}, \
  and\ \bibinfo {author} {\bibfnamefont {I.}~\bibnamefont {Dodin}},\ }\href
  {\doibase 10.1017/S0022377818001307} {\bibfield  {journal} {\bibinfo
  {journal} {Journal of Plasma Physics}\ }\textbf {\bibinfo {volume} {85}},\
  \bibinfo {pages} {1} (\bibinfo {year} {2019})}\BibitemShut {NoStop}%
\bibitem [{\citenamefont {Zhu}\ and\ \citenamefont {Dodin}(2021)}]{ZhuDodin}%
  \BibitemOpen
  \bibfield  {author} {\bibinfo {author} {\bibfnamefont {H.}~\bibnamefont
  {Zhu}}\ and\ \bibinfo {author} {\bibfnamefont {I.~Y.}\ \bibnamefont
  {Dodin}},\ }\href {https://api.semanticscholar.org/CorpusID:231583190}
  {\bibfield  {journal} {\bibinfo  {journal} {Physics of Plasmas}\ }\textbf
  {\bibinfo {volume} {28}},\ \bibinfo {pages} {032303} (\bibinfo {year}
  {2021})}\BibitemShut {NoStop}%
\bibitem [{\citenamefont {Vallis}(2006)}]{Va}%
  \BibitemOpen
  \bibfield  {author} {\bibinfo {author} {\bibfnamefont {G.~K.}\ \bibnamefont
  {Vallis}},\ }\href@noop {} {\emph {\bibinfo {title} {Atmospheric and Oceanic
  Fluid Dynamics. Fundamentals and Large-scale Circulation}}}\ (\bibinfo
  {publisher} {Cambridge},\ \bibinfo {year} {2006})\BibitemShut {NoStop}%
\bibitem [{\citenamefont {Hide}(1966)}]{Hide66}%
  \BibitemOpen
  \bibfield  {author} {\bibinfo {author} {\bibfnamefont {R.}~\bibnamefont
  {Hide}},\ }\href@noop {} {\bibfield  {journal} {\bibinfo  {journal} {Proc.
  Roy. Soc. London}\ }\textbf {\bibinfo {volume} {259}},\ \bibinfo {pages}
  {615} (\bibinfo {year} {1966})}\BibitemShut {NoStop}%
\bibitem [{\citenamefont {Zaqarashvili}\ \emph {et~al.}(2007)\citenamefont
  {Zaqarashvili}, \citenamefont {Oliver}, \citenamefont {Ballester},\ and\
  \citenamefont {Shergelashvili}}]{ZaqOliBalShe}%
  \BibitemOpen
  \bibfield  {author} {\bibinfo {author} {\bibfnamefont {T.~V.}\ \bibnamefont
  {Zaqarashvili}}, \bibinfo {author} {\bibfnamefont {R.}~\bibnamefont
  {Oliver}}, \bibinfo {author} {\bibfnamefont {J.~L.}\ \bibnamefont
  {Ballester}}, \ and\ \bibinfo {author} {\bibfnamefont {B.~M.}\ \bibnamefont
  {Shergelashvili}},\ }\href@noop {} {\bibfield  {journal} {\bibinfo  {journal}
  {A\&A}\ }\textbf {\bibinfo {volume} {470}},\ \bibinfo {pages} {815} (\bibinfo
  {year} {2007})}\BibitemShut {NoStop}%
\bibitem [{\citenamefont {Balk}(2014)}]{Balk2014}%
  \BibitemOpen
  \bibfield  {author} {\bibinfo {author} {\bibfnamefont {A.~M.}\ \bibnamefont
  {Balk}},\ }\href@noop {} {\bibfield  {journal} {\bibinfo  {journal} {ApJ}\
  }\textbf {\bibinfo {volume} {796}},\ \bibinfo {pages} {143 (8pp)} (\bibinfo
  {year} {2014})}\BibitemShut {NoStop}%
\bibitem [{\citenamefont {Balk}(2022)}]{B2022}%
  \BibitemOpen
  \bibfield  {author} {\bibinfo {author} {\bibfnamefont {A.~M.}\ \bibnamefont
  {Balk}},\ }\href {\doibase 10.3847/1538-4357/ac448d} {\bibfield  {journal}
  {\bibinfo  {journal} {The Astrophysical Journal}\ }\textbf {\bibinfo {volume}
  {926}},\ \bibinfo {pages} {2} (\bibinfo {year} {2022})}\BibitemShut {NoStop}%
\bibitem [{\citenamefont {Braginsky}(1998)}]{Brag98}%
  \BibitemOpen
  \bibfield  {author} {\bibinfo {author} {\bibfnamefont {S.~I.}\ \bibnamefont
  {Braginsky}},\ }\href@noop {} {\bibfield  {journal} {\bibinfo  {journal}
  {Earth Planets Space}\ }\textbf {\bibinfo {volume} {50}},\ \bibinfo {pages}
  {641} (\bibinfo {year} {1998})}\BibitemShut {NoStop}%
\bibitem [{\citenamefont {Boltzmann}(1875)}]{Boltz1}%
  \BibitemOpen
  \bibfield  {author} {\bibinfo {author} {\bibfnamefont {L.}~\bibnamefont
  {Boltzmann}},\ }\href@noop {} {\bibfield  {journal} {\bibinfo  {journal}
  {Wien. Ber.}\ }\textbf {\bibinfo {volume} {72}},\ \bibinfo {pages} {427}
  (\bibinfo {year} {1875})}\BibitemShut {NoStop}%
\bibitem [{\citenamefont {Boltzmann}(1876)}]{Boltz2}%
  \BibitemOpen
  \bibfield  {author} {\bibinfo {author} {\bibfnamefont {L.}~\bibnamefont
  {Boltzmann}},\ }\href@noop {} {\bibfield  {journal} {\bibinfo  {journal}
  {Sitzungsberichte der academie der wis\-sen\-schaften wien}\ }\textbf
  {\bibinfo {volume} {74}},\ \bibinfo {pages} {503} (\bibinfo {year}
  {1876})}\BibitemShut {NoStop}%
\bibitem [{\citenamefont {Cercignani}(1990)}]{Boltzmann}%
  \BibitemOpen
  \bibfield  {author} {\bibinfo {author} {\bibfnamefont {C.}~\bibnamefont
  {Cercignani}},\ }\href@noop {} {\bibfield  {journal} {\bibinfo  {journal} {J.
  Stat. Phys.}\ }\textbf {\bibinfo {volume} {58}},\ \bibinfo {pages} {817}
  (\bibinfo {year} {1990})}\BibitemShut {NoStop}%
\bibitem [{\citenamefont {Balk}\ and\ \citenamefont {Ferapontov}(1998)}]{BaFe}%
  \BibitemOpen
  \bibfield  {author} {\bibinfo {author} {\bibfnamefont {A.~M.}\ \bibnamefont
  {Balk}}\ and\ \bibinfo {author} {\bibfnamefont {E.~V.}\ \bibnamefont
  {Ferapontov}},\ }in\ \href@noop {} {\emph {\bibinfo {booktitle} {Nonlinear
  waves and weak turbulence}}},\ \bibinfo {editor} {edited by\ \bibinfo
  {editor} {\bibfnamefont {V.~E.}\ \bibnamefont {Zakharov}}}\ (\bibinfo
  {publisher} {Amer.\ Math.\ Soc.\ Trans.\ Ser. 2, vol. 182},\ \bibinfo {year}
  {1998})\ pp.\ \bibinfo {pages} {1--30}\BibitemShut {NoStop}%
\bibitem [{\citenamefont {Balk}\ \emph {et~al.}(1991)\citenamefont {Balk},
  \citenamefont {Nazarenko},\ and\ \citenamefont {Zakharov}}]{BNZ}%
  \BibitemOpen
  \bibfield  {author} {\bibinfo {author} {\bibfnamefont {A.~M.}\ \bibnamefont
  {Balk}}, \bibinfo {author} {\bibfnamefont {S.~V.}\ \bibnamefont {Nazarenko}},
  \ and\ \bibinfo {author} {\bibfnamefont {V.~E.}\ \bibnamefont {Zakharov}},\
  }\href@noop {} {\bibfield  {journal} {\bibinfo  {journal} {Phys.\ Lett.\ A}\
  }\textbf {\bibinfo {volume} {152}},\ \bibinfo {pages} {276} (\bibinfo {year}
  {1991})}\BibitemShut {NoStop}%
\bibitem [{\citenamefont {Balk}(1991)}]{B1991}%
  \BibitemOpen
  \bibfield  {author} {\bibinfo {author} {\bibfnamefont {A.~M.}\ \bibnamefont
  {Balk}},\ }\href@noop {} {\bibfield  {journal} {\bibinfo  {journal} {Phys.\
  Lett.\ A}\ }\textbf {\bibinfo {volume} {155}},\ \bibinfo {pages} {20}
  (\bibinfo {year} {1991})}\BibitemShut {NoStop}%
\bibitem [{\citenamefont {Zakharov}\ and\ \citenamefont
  {Schulman}(1980)}]{ZSch0}%
  \BibitemOpen
  \bibfield  {author} {\bibinfo {author} {\bibfnamefont {V.~E.}\ \bibnamefont
  {Zakharov}}\ and\ \bibinfo {author} {\bibfnamefont {E.~I.}\ \bibnamefont
  {Schulman}},\ }\href@noop {} {\bibfield  {journal} {\bibinfo  {journal}
  {Physica D}\ }\textbf {\bibinfo {volume} {1}},\ \bibinfo {pages} {192}
  (\bibinfo {year} {1980})}\BibitemShut {NoStop}%
\bibitem [{\citenamefont {Zakharov}\ and\ \citenamefont
  {Schulman}(1988)}]{ZSch}%
  \BibitemOpen
  \bibfield  {author} {\bibinfo {author} {\bibfnamefont {V.~E.}\ \bibnamefont
  {Zakharov}}\ and\ \bibinfo {author} {\bibfnamefont {E.~I.}\ \bibnamefont
  {Schulman}},\ }\href@noop {} {\bibfield  {journal} {\bibinfo  {journal}
  {Physica D}\ }\textbf {\bibinfo {volume} {29}},\ \bibinfo {pages} {283}
  (\bibinfo {year} {1988})}\BibitemShut {NoStop}%
\bibitem [{\citenamefont {Ferapontov}(1992)}]{Ferapontov}%
  \BibitemOpen
  \bibfield  {author} {\bibinfo {author} {\bibfnamefont {E.~V.}\ \bibnamefont
  {Ferapontov}},\ }\enquote {\bibinfo {title} {Web geometry and mathematical
  physics},}\ \ (\bibinfo  {publisher} {Kluwer Academic Publishers, Norwell,
  MA},\ \bibinfo {year} {1992})\ pp.\ \bibinfo {pages} {310--323},\ \bibinfo
  {note} {appendix in the book ``{G}eometry and {A}lgebra of {M}ultidimensional
  {T}hree-{W}ebs'' by {M}. {A}. {A}kivis and {A}. {M}. {S}helekhov}\BibitemShut
  {NoStop}%
\bibitem [{\citenamefont {Blaschke}(1932)}]{Bl1}%
  \BibitemOpen
  \bibfield  {author} {\bibinfo {author} {\bibfnamefont {W.}~\bibnamefont
  {Blaschke}},\ }\href@noop {} {\emph {\bibinfo {title} {Topological
  differential geometry}}}\ (\bibinfo  {publisher} {Chicago University Press},\
  \bibinfo {year} {1932})\BibitemShut {NoStop}%
\bibitem [{\citenamefont {Balk}(2005)}]{B2005}%
  \BibitemOpen
  \bibfield  {author} {\bibinfo {author} {\bibfnamefont {A.~M.}\ \bibnamefont
  {Balk}},\ }\href@noop {} {\bibfield  {journal} {\bibinfo  {journal} {Phys.\
  Lett.\ A}\ }\textbf {\bibinfo {volume} {345}},\ \bibinfo {pages} {154}
  (\bibinfo {year} {2005})}\BibitemShut {NoStop}%
\bibitem [{\citenamefont {Galperin}\ \emph {et~al.}(2006)\citenamefont
  {Galperin}, \citenamefont {Sukoriansky}, \citenamefont {Dikovskaya},
  \citenamefont {Read}, \citenamefont {Yamazaki},\ and\ \citenamefont
  {Words\-worth}}]{Glp}%
  \BibitemOpen
  \bibfield  {author} {\bibinfo {author} {\bibfnamefont {B.}~\bibnamefont
  {Galperin}}, \bibinfo {author} {\bibfnamefont {S.}~\bibnamefont
  {Sukoriansky}}, \bibinfo {author} {\bibfnamefont {N.}~\bibnamefont
  {Dikovskaya}}, \bibinfo {author} {\bibfnamefont {P.~L.}\ \bibnamefont
  {Read}}, \bibinfo {author} {\bibfnamefont {Y.~H.}\ \bibnamefont {Yamazaki}},
  \ and\ \bibinfo {author} {\bibfnamefont {R.}~\bibnamefont {Words\-worth}},\
  }\href@noop {} {\bibfield  {journal} {\bibinfo  {journal} {Nonlin. Processes
  Geophys.}\ }\textbf {\bibinfo {volume} {13}},\ \bibinfo {pages} {89}
  (\bibinfo {year} {2006})}\BibitemShut {NoStop}%
\bibitem [{\citenamefont {Weichman}(2006)}]{Weichman06}%
  \BibitemOpen
  \bibfield  {author} {\bibinfo {author} {\bibfnamefont {P.}~\bibnamefont
  {Weichman}},\ }\href {\doibase 10.1103/PhysRevE.73.036313} {\bibfield
  {journal} {\bibinfo  {journal} {Physical review. E, Statistical, nonlinear,
  and soft matter physics}\ }\textbf {\bibinfo {volume} {73}},\ \bibinfo
  {pages} {036313} (\bibinfo {year} {2006})}\BibitemShut {NoStop}%
\bibitem [{\citenamefont {Balk}\ and\ \citenamefont {van
  Heerden}(2006)}]{BaVan}%
  \BibitemOpen
  \bibfield  {author} {\bibinfo {author} {\bibfnamefont {A.~M.}\ \bibnamefont
  {Balk}}\ and\ \bibinfo {author} {\bibfnamefont {F.}~\bibnamefont {van
  Heerden}},\ }\href@noop {} {\bibfield  {journal} {\bibinfo  {journal}
  {Physica D}\ }\textbf {\bibinfo {volume} {223}},\ \bibinfo {pages} {109}
  (\bibinfo {year} {2006})}\BibitemShut {NoStop}%
\bibitem [{\citenamefont {Sukoriansky}\ \emph {et~al.}(2007)\citenamefont
  {Sukoriansky}, \citenamefont {Dikovskaya},\ and\ \citenamefont
  {Galperin}}]{SukorianskyDG}%
  \BibitemOpen
  \bibfield  {author} {\bibinfo {author} {\bibfnamefont {S.}~\bibnamefont
  {Sukoriansky}}, \bibinfo {author} {\bibfnamefont {N.}~\bibnamefont
  {Dikovskaya}}, \ and\ \bibinfo {author} {\bibfnamefont {B.}~\bibnamefont
  {Galperin}},\ }\href {\doibase https://doi.org/10.1175/JAS4013.1} {\bibfield
  {journal} {\bibinfo  {journal} {Journal of the Atmospheric Sciences}\
  }\textbf {\bibinfo {volume} {64}},\ \bibinfo {pages} {3312 } (\bibinfo {year}
  {2007})}\BibitemShut {NoStop}%
\bibitem [{\citenamefont {Lee}\ and\ \citenamefont {Smith}(2007)}]{LeeSmith}%
  \BibitemOpen
  \bibfield  {author} {\bibinfo {author} {\bibfnamefont {Y.}~\bibnamefont
  {Lee}}\ and\ \bibinfo {author} {\bibfnamefont {L.~M.}\ \bibnamefont
  {Smith}},\ }\href@noop {} {\bibfield  {journal} {\bibinfo  {journal} {J.
  Fluid Mech.}\ }\textbf {\bibinfo {volume} {576}},\ \bibinfo {pages} {405}
  (\bibinfo {year} {2007})}\BibitemShut {NoStop}%
\bibitem [{\citenamefont {Read}\ \emph {et~al.}(2007)\citenamefont {Read},
  \citenamefont {Yamazaki}, \citenamefont {Lewis}, \citenamefont {Williams},
  \citenamefont {Wordsworth}, \citenamefont {Miki-Yamazaki}, \citenamefont
  {Sommeria},\ and\ \citenamefont {Didelle}}]{Read+6}%
  \BibitemOpen
  \bibfield  {author} {\bibinfo {author} {\bibfnamefont {P.~L.}\ \bibnamefont
  {Read}}, \bibinfo {author} {\bibfnamefont {Y.~H.}\ \bibnamefont {Yamazaki}},
  \bibinfo {author} {\bibfnamefont {S.~R.}\ \bibnamefont {Lewis}}, \bibinfo
  {author} {\bibfnamefont {P.~D.}\ \bibnamefont {Williams}}, \bibinfo {author}
  {\bibfnamefont {R.~D.}\ \bibnamefont {Wordsworth}}, \bibinfo {author}
  {\bibfnamefont {K.}~\bibnamefont {Miki-Yamazaki}}, \bibinfo {author}
  {\bibfnamefont {J.}~\bibnamefont {Sommeria}}, \ and\ \bibinfo {author}
  {\bibfnamefont {H.}~\bibnamefont {Didelle}},\ }\href
  {https://api.semanticscholar.org/CorpusID:15945084} {\bibfield  {journal}
  {\bibinfo  {journal} {Journal of the Atmospheric Sciences}\ }\textbf
  {\bibinfo {volume} {64}},\ \bibinfo {pages} {4031} (\bibinfo {year}
  {2007})}\BibitemShut {NoStop}%
\bibitem [{\citenamefont {Yoshi-Yuki}\ \emph {et~al.}(2007)\citenamefont
  {Yoshi-Yuki}, \citenamefont {Nishizawa}, \citenamefont {Takehiro},
  \citenamefont {Yamada}, \citenamefont {Ishioka},\ and\ \citenamefont
  {Yoden}}]{Yoshi-Yuki07}%
  \BibitemOpen
  \bibfield  {author} {\bibinfo {author} {\bibfnamefont {H.}~\bibnamefont
  {Yoshi-Yuki}}, \bibinfo {author} {\bibfnamefont {S.}~\bibnamefont
  {Nishizawa}}, \bibinfo {author} {\bibfnamefont {S.-I.}\ \bibnamefont
  {Takehiro}}, \bibinfo {author} {\bibfnamefont {M.}~\bibnamefont {Yamada}},
  \bibinfo {author} {\bibfnamefont {K.}~\bibnamefont {Ishioka}}, \ and\
  \bibinfo {author} {\bibfnamefont {S.}~\bibnamefont {Yoden}},\ }\href
  {\doibase 10.1175/2007JAS2209.1} {\bibfield  {journal} {\bibinfo  {journal}
  {Journal of The Atmospheric Sciences - J ATMOS SCI}\ }\textbf {\bibinfo
  {volume} {64}},\ \bibinfo {pages} {4246} (\bibinfo {year}
  {2007})}\BibitemShut {NoStop}%
\bibitem [{\citenamefont {Balk}\ and\ \citenamefont {Yoshikawa}(2009)}]{BaYo}%
  \BibitemOpen
  \bibfield  {author} {\bibinfo {author} {\bibfnamefont {A.~M.}\ \bibnamefont
  {Balk}}\ and\ \bibinfo {author} {\bibfnamefont {T.}~\bibnamefont
  {Yoshikawa}},\ }\href {\doibase 10.1016/j.physd.2008.11.008} {\bibfield
  {journal} {\bibinfo  {journal} {Physica D}\ }\textbf {\bibinfo {volume}
  {238}},\ \bibinfo {pages} {384} (\bibinfo {year} {2009})}\BibitemShut
  {NoStop}%
\bibitem [{\citenamefont {Nazarenko}\ and\ \citenamefont
  {Quinn}(2009)}]{Nazarenko}%
  \BibitemOpen
  \bibfield  {author} {\bibinfo {author} {\bibfnamefont {S.}~\bibnamefont
  {Nazarenko}}\ and\ \bibinfo {author} {\bibfnamefont {B.}~\bibnamefont
  {Quinn}},\ }\href@noop {} {\bibfield  {journal} {\bibinfo  {journal} {Phys.
  Rev. Lett.}\ }\textbf {\bibinfo {volume} {103}},\ \bibinfo {pages} {118501}
  (\bibinfo {year} {2009})}\BibitemShut {NoStop}%
\bibitem [{\citenamefont {Balk}\ and\ \citenamefont
  {Zakharov}(2009)}]{BaZaSimilarity}%
  \BibitemOpen
  \bibfield  {author} {\bibinfo {author} {\bibfnamefont {A.~M.}\ \bibnamefont
  {Balk}}\ and\ \bibinfo {author} {\bibfnamefont {V.~E.}\ \bibnamefont
  {Zakharov}},\ }\href@noop {} {\bibfield  {journal} {\bibinfo  {journal}
  {Phys. Lett. A}\ }\textbf {\bibinfo {volume} {373}},\ \bibinfo {pages} {4049}
  (\bibinfo {year} {2009})}\BibitemShut {NoStop}%
\bibitem [{\citenamefont {Balk}\ \emph {et~al.}(2011)\citenamefont {Balk},
  \citenamefont {van Heerden},\ and\ \citenamefont {Weichman}}]{BHW}%
  \BibitemOpen
  \bibfield  {author} {\bibinfo {author} {\bibfnamefont {A.~M.}\ \bibnamefont
  {Balk}}, \bibinfo {author} {\bibfnamefont {F.}~\bibnamefont {van Heerden}}, \
  and\ \bibinfo {author} {\bibfnamefont {P.~B.}\ \bibnamefont {Weichman}},\
  }\href@noop {} {\bibfield  {journal} {\bibinfo  {journal} {Phys. Rev. E}\
  }\textbf {\bibinfo {volume} {83}},\ \bibinfo {pages} {046320} (\bibinfo
  {year} {2011})}\BibitemShut {NoStop}%
\bibitem [{\citenamefont {Weichman}(2012)}]{Weichman12}%
  \BibitemOpen
  \bibfield  {author} {\bibinfo {author} {\bibfnamefont {P.}~\bibnamefont
  {Weichman}},\ }\href {\doibase 10.1103/PhysRevLett.109.235002} {\bibfield
  {journal} {\bibinfo  {journal} {Physical review letters}\ }\textbf {\bibinfo
  {volume} {109}},\ \bibinfo {pages} {235002} (\bibinfo {year}
  {2012})}\BibitemShut {NoStop}%
\bibitem [{\citenamefont {Saito}\ and\ \citenamefont
  {Ishioka}(2013)}]{SaitoIshioka13}%
  \BibitemOpen
  \bibfield  {author} {\bibinfo {author} {\bibfnamefont {I.}~\bibnamefont
  {Saito}}\ and\ \bibinfo {author} {\bibfnamefont {K.}~\bibnamefont
  {Ishioka}},\ }\href@noop {} {\bibfield  {journal} {\bibinfo  {journal} {Phys.
  Fluids}\ }\textbf {\bibinfo {volume} {25}},\ \bibinfo {pages} {076602}
  (\bibinfo {year} {2013})}\BibitemShut {NoStop}%
\bibitem [{\citenamefont {Saito}\ and\ \citenamefont
  {Ishioka}(2016)}]{SaitoIshioka16}%
  \BibitemOpen
  \bibfield  {author} {\bibinfo {author} {\bibfnamefont {I.}~\bibnamefont
  {Saito}}\ and\ \bibinfo {author} {\bibfnamefont {K.}~\bibnamefont
  {Ishioka}},\ }\href {\doibase 10.2151/jmsj.2016-002} {\bibfield  {journal}
  {\bibinfo  {journal} {J. Met. Soc. Japan. Ser. II}\ }\textbf {\bibinfo
  {volume} {94}},\ \bibinfo {pages} {25} (\bibinfo {year} {2016})}\BibitemShut
  {NoStop}%
\bibitem [{\citenamefont {Zakharov}(1985)}]{ZakhKS}%
  \BibitemOpen
  \bibfield  {author} {\bibinfo {author} {\bibfnamefont {V.~E.}\ \bibnamefont
  {Zakharov}},\ }in\ \href@noop {} {\emph {\bibinfo {booktitle} {Basic plasma
  physics}}},\ Vol.~\bibinfo {volume} {2},\ \bibinfo {editor} {edited by\
  \bibinfo {editor} {\bibfnamefont {A.~A.}\ \bibnamefont {Galeev}}\ and\
  \bibinfo {editor} {\bibfnamefont {R.~N.}\ \bibnamefont {Sudan}}}\ (\bibinfo
  {publisher} {Elsevier (North-Holland Publishing Company), Amsterdam},\
  \bibinfo {year} {1985})\ pp.\ \bibinfo {pages} {3--36}\BibitemShut {NoStop}%
\bibitem [{\citenamefont {Balk}(2019)}]{B19}%
  \BibitemOpen
  \bibfield  {author} {\bibinfo {author} {\bibfnamefont {A.~M.}\ \bibnamefont
  {Balk}},\ }\href {\doibase 10.1103/PhysRevResearch.1.033180} {\bibfield
  {journal} {\bibinfo  {journal} {Phys. Rev. Res.}\ }\textbf {\bibinfo {volume}
  {1}},\ \bibinfo {pages} {033180} (\bibinfo {year} {2019})}\BibitemShut
  {NoStop}%
\end{thebibliography}%
\end{document}